\documentclass[trackchanges]{aastex701}

\usepackage{natbib}

\begin{document}

\title{Wave-particle acceleration during supernova explosions}

\author[orcid=0000-0002-9657-5356]{Z.N. Osmanov}
\affiliation{School of Physics, Free University of Tbilisi, 0183, Tbilisi, Georgia}
\affiliation{E. Kharadze Georgian National Astrophysical Observatory, Abastumani 0301, Georgia}
\email[show]{z.osmanov@freeuni.edu.ge}  

\author[orcid=]{S.M. Majajan} 
\affiliation{Institute for Fusion Studies, The University of Texas at Austin, Austin, Texas 78712}
\email{fakeemail2@google.com}


\begin{abstract}

We investigate particle acceleration during the early stages of a supernova explosion via a resonant interaction between electromagnetic waves and quantum waves associated with  particles. For typical progenitor masses of $(8; 12; 20)\times M_s$, we find that this mechanism can accelerate protons to energies of up to $3\times 10^{14}$ eV. Interestingly, despite the continuously declining luminosity, the particle energy initially increases, reaches a maximum, and subsequently decreases, reflecting the characteristic behavior of the resonant acceleration mechanism.

\end{abstract}



\section{Introduction}

The origin of very high energy (VHE) cosmic rays (CR) continues to remain one of the challenging fundamental problems in modern astrophysics. There are two distinct questions that need to be answered:1) Where - What astrophysical/cosmological settings are suitable for such extreme energization, and 2) How - What physical mechanisms could exploit such conditions to produce VHE
 particles. 
 
 For the former, two arenas have been studied in some detail; 1)the magnetospheres of supermassive black holes or pulsars \cite{agn,NO,pulsars}, and 2)astrophysical settings such as supernova explosions. \cite{SN1,SN2,SN3}.

Regarding the particle energization, two most invoked mechanisms are: the  magneto-centrifugal\cite{review} and Fermi acceleration \cite{fermi} (and their various modifications \cite{SN1,bell}.

In this paper, we pursue a recently proposed mechanism, based on a resonant wave-wave energy transfer. The essence as well as the novelty of the approach lies in viewing the erstwhile particle as a quantum wave described by a Klein-Gordon (or Dirac) equation. For simplicity, we will limit ourselves to dealing with the spinless Klein Gordon particle (KGP)wave. For this high energy exchange, it is the relativistic KGP that is drawing energy from a back ground high intensity E.M wave (a similar calculation could be done, for instance, for intense gravitational waves). 

The basic coceptual framework for KGP-EM wave-wave interaction was developed in \cite{MA1,MA2}. Since then, it has been successfully employed to VHE production 
in the environments of black holes, pulsars, and the solar atmosphere. \cite{res1,res2,res3}.

We will now examine if the electromagnetic radiation generated during a supernova explosion could resonantly accelerate KGP waves.  

The paper is organized as follows. In Section 2, after summarizing the relevant physics behind the resonant wave-wave mechanism, we will work out how it will operate in the aftermath of a supernova explosion. In Section 3, we summarize the main results derived in Section2.

\section{Resonant Wave-Wave Interaction and a supernova explosion}

In this section, we discuss the phenomenon of resonant particle acceleration and investigate the temporal evolution of the maximum attainable particle energies during a supernova explosion. 

\subsection{Resonance acceleration mechanism}
Following the original papers \cite{MA1,MA2}, we summarize here the relevant parts of the general theory. We start by noting that  the group velocity of a relativistic KGP 
\begin{equation}
\label{vg} 
\upsilon_g = \frac{\partial E}{\partial P} = \frac{P}{\left(P^2+m^2\right)^{1/2}},
\end{equation}
approaches the speed of light ($c = 1$)at high energies, when($p>>m$); the quantum wave associated with such KGP can, thus, come into resonance with the electromagnetic radiation and accelerate drawing energy from the E.M wave. Here, $E$ and $P$, $m$ represent, respectively, the energy, momentum and mass of the KGP.

 We consider an electromagnetic (EM) wave propagating along the z-axis and, for simplicity, assume it to be circularly polarized: $A^0 = A^z = 0$, $A^x = A\cos\left(\omega t-kz\right)$ and $A^y = -A\cos\left(\omega t-kz\right)$. Then, the wave function, $\Psi$ of the KGP, obeys \cite{MA1}
\begin{equation}\label{KG} 
\left(\partial^2_t-\partial_z^2+2eAK_{\perp}\cos\left(\omega t-kz\right)\right)\Psi+\left(K_{\perp}^2+m^2+e^2A^2\right)\Psi = 0,
\end{equation}
where $e$ is the particle's charge, $\omega$, ($k$) is the frequency (wave number) of the EM radiation, and $K_{\perp}$ represents the perpendicular momentum of the KGP in the (ignorable)direction perpendicular to that of the EM wave propagation.  

Following standard methods, (\ref{KG}) can be analyzed as an ordinary differential equation (o.d.e) in the variable $\xi=\omega t - kz$. The resulting Mathew equation
\begin{equation}
\label{mathew} 
\left(\omega^2-k^2\right)\frac{d^2\psi}{d\xi^2}+\left(\mu+\nu\cos\xi\right) = 0,
\end{equation}
with $\mu = K_{\perp}^2+m^2+e^2A^2$ and $\nu = 2eAK_{\perp}$, is clearly singular  when $\omega^2-k^2$ approaches zero. In this limit, ${d^2\psi}/{d\xi^2}$ must become large for the equation to be satisfied. Thus, in this limit, the KGP energy -momentum (proportional to the derivative $\psi$) will be resonantly enhanced. 

Although(\ref{mathew}) allows exact analysis, we can extract most relevant information by seeking a WKB solution (which becomes better and better as $\omega^2-k^2$ approaches zero). Referring the reader to \cite{MA2}for details, we quote here the most relevant result- the resonant acceleration, after using the diaprsion relation ($\omega^2-c^2k^2 = {\omega_p^2}\times\left({1+e^2A^2/(m^2c^4)}\right)^{-1/2}$), comes out to be 

\begin{equation}
\label{dEdt} 
\frac{dE}{dt}\simeq ecAK_{\perp}\frac{\sqrt{2}\hbar\omega^2}{\omega_p}\frac{\left(1+\frac{e^2A^2}{m^2c^4}\right)^{1/4}}{\left(1+\frac{e^2A^2}{m^2c^4}+\frac{\hbar^2K_{\perp}^2c^2}{m^2c^4}\right)^{1/2}},
\end{equation}
where natural dimensions of the physical variable are restored. Here $\hbar$ is the Planck's constant, $\omega_p = \sqrt{4\pi ne^2/m_e}$ is the plasma frequency, $n$ denotes the number density and $m_e$ is the electron's mass. Since $\omega_p = \sqrt{4\pi ne^2/m_e}$ is the measure of the difference $\omega^2-c^2k^2$,  $\omega_p<<\omega$ near resonance.

As demonstrated in \cite{MA2}, the above expression yields the maximum energy attainable by a particle, which is given by
\begin{equation}
\label{E} 
E_{max}\simeq mc^2\times \frac{\omega}{\omega_p}\times\left(1+\frac{e^2A^2}{m^2c^4}\right)^{1/4}\times\left(1+\frac{e^2A^2}{m^2c^4}+\frac{\hbar^2K_{\perp}^2c^2}{m^2c^4}\right)^{1/2}.
\end{equation}
\subsection{Application to supernova explosions}
What can we transport  Equation (\ref{E}) to a supernove explosion? 

Since our aim is not to investigate the details of the supernova explosion itself but rather to study particle acceleration, we adopt a simple toy model for the explosion itself. A progenitor star of mass $M$, after the explosion leaves behind a neutron star with a mass of $1.5 M_s$, where $M_s\simeq 2\times 10^{33}$g is the solar mass. The ejected mass, $M_{ej} = M-1.5 M_s$ corresponding to an explosion energy of $E_{SN}\simeq 10^{51}$ ergs \cite{carroll}, will acquire a characteristic velocity 
\begin{equation}
\label{vel} 
\upsilon\simeq\left(\frac{2E_{SN}}{M_{ej}}\right)^{1/2}\simeq 3200\times\left(\frac{E_{51}}{M_{10}-0.15}\right)^{1/2}\; km/s,
\end{equation}
where $E_{51} = E_{SN}/(10^{51}erg)$ and $M_{10} = 10M_s$. This velocity remains approximately constant until the mass of the swept-up interstellar medium (ISM) becomes comparable to that of the ejecta. It is straightforward to verify that, for the velocity estimated above, the mass of the accumulated ISM after one year is approximately nine orders of magnitude smaller than the solar mass. Therefore, over the time interval considered in our calculations, we can safely assume that the ejecta velocity remains constant.

\begin{figure}
  \centering {\includegraphics[width=12cm]{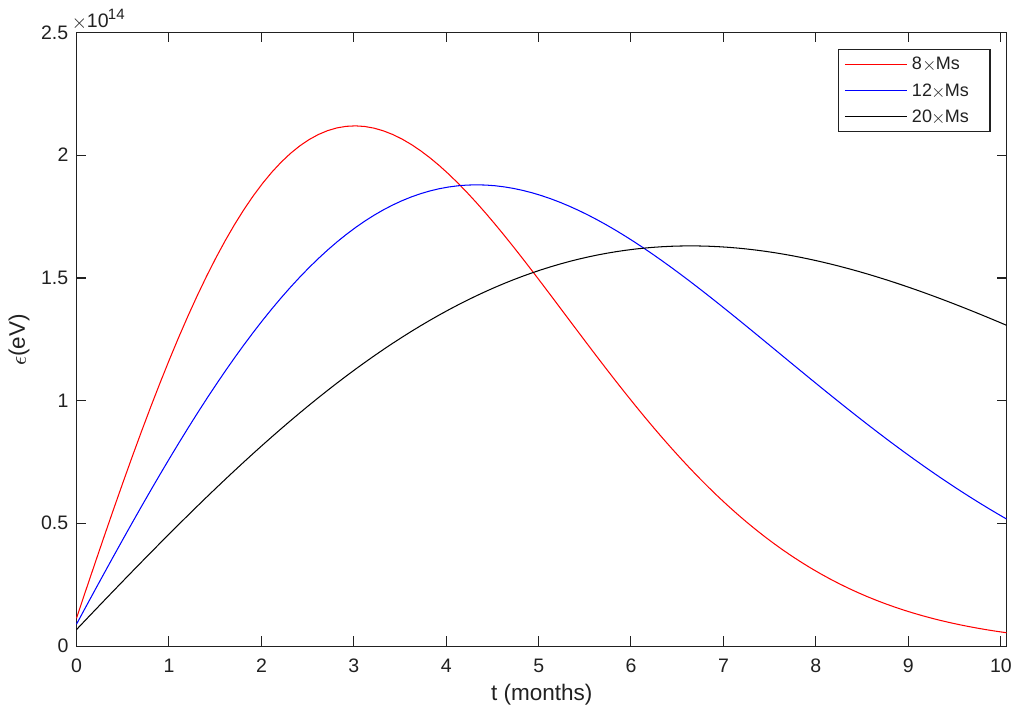}}
 \caption{Time evolution of maximum attainable energies of protons for three different progenitor masses $[8; 12; 20]\times M_s$. The set of parameters is: $L_0 = 10^{43}$ erg/s; $R_0 = 10^{14}$ cm, $\kappa = 0.1$ \cite{lumin}.}\label{fig1}
\end{figure}

\begin{figure}
  \centering {\includegraphics[width=12cm]{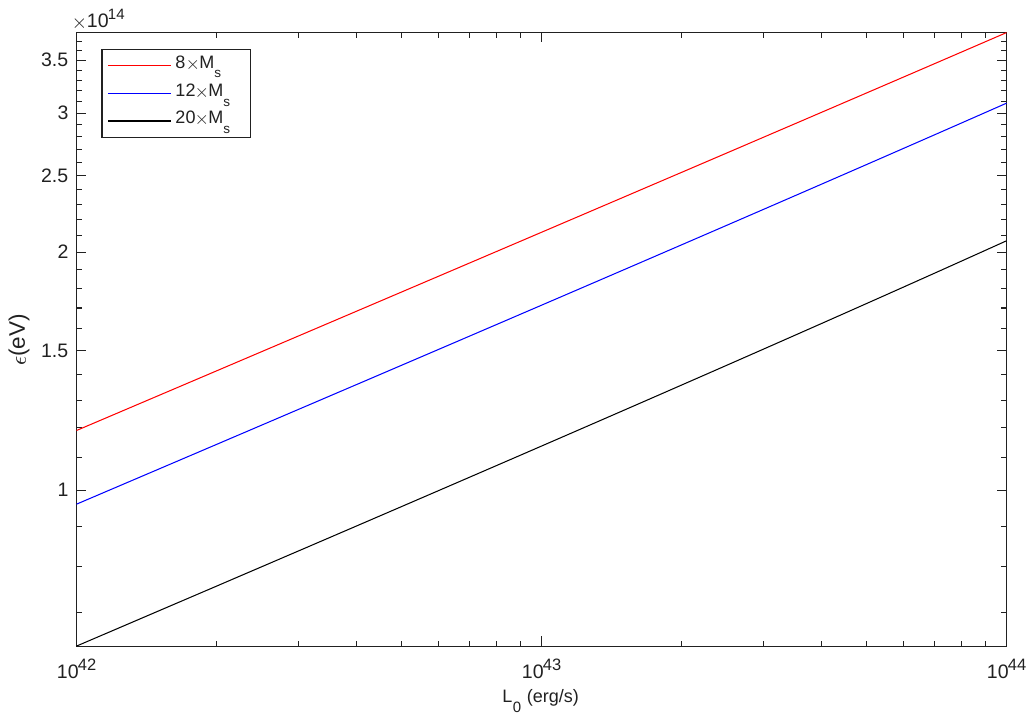}}
 \caption{The behaviour of $\epsilon$ versus $L_0$. The set of parameters is the same as in the previous case, except: $t = 3$ months and $L = [10^{42}; 10^{44}] erg/s$.}\label{fig2}
\end{figure}

During the expansion, the luminosity gradually decreases from its peak value \cite{lumin}
\begin{equation}
\label{lumin} 
L\simeq L_0\exp{\left(-\frac{\left(2R_0+\upsilon t\right)t}{2k}\right)},
\end{equation}
where $L_0$ represents the maximum luminosity during the explosion, $R_0$ is the corresponding radius of the supernova bubble, $k = 3\kappa M/(4\pi c)$ and $\kappa$ is the opacity of a medium. 

As a first approximation, following \cite{lumin}, we assume that the radiation can be described as blackbody radiation, with the corresponding radiation temperature determined by the Planck law
\begin{equation}
\label{T} 
T\simeq\left(\frac{L}{4\pi\sigma (R_0+\upsilon t)^2}\right)^{1/4},
\end{equation}
where $\sigma$ represents the Stefan-Boltzmann constant. It then follows that the wavelength corresponding to the maximum of the spectral flux is determined by Wien’s displacement law,
$\lambda_{\max} = b/T$.
Through Eq. (\ref{T}), this wavelength determines the corresponding  frequency $\omega = 2\pi c/\lambda$. A particle that comes into resonance at this frequency will therefore attain the maximum possible energy (see Eq. (\ref{E})).

Taking into account $A = c\sqrt{cL/\pi}/(2\omega(R_0+\upsilon t))$ \cite{res3}, and assuming that the density of particles in the ejecta can be approximated as follows ($m_p$ is the proton's mass)
\begin{equation}
\label{n} 
n\simeq\frac{3M_{ej}}{4\pi m_p\left(R_0+\upsilon t\right)^3},
\end{equation}
one can readily estimate the maximum energies.

On Fig. 1 we show the behaviour of maximum attainable energies of protons versus time for three different progenitor masses $(8; 12; 20)\times M_s$. The set of parameters is: $L_0 = 10^{43}$ erg/s; $R_0 = 10^{14}$ cm, $\kappa = 0.1$ \cite{lumin}. As it is clear, protons may reach enormous energies of the order of hundreds of TeVs.

Fig. 2 shows the same quantities as a function of the initial luminosity. The parameter set is the same, except for the fixed time $t = 3$ months and the luminosity interval. As expected, an increase in the initial luminosity leads to a monotonic increase in the energy imparted to the protons through acceleration.

Although the luminosity decreases monotonically with time, the energy follows a different trajectory; it increases initially, reaches a maximum, and subsequently begins to decrease as well. This behavior can be understood from the relation $\omega/\omega_p\propto (R_0+\upsilon t) L^{1/4}$. At early times, the radius $R_0+\upsilon t$ increases more rapidly than $L^{1/4}$ decreases, causing the particle energy—to increase. As the expansion proceeds, however, the decline in luminosity becomes sufficiently rapid to overcome the increase in $R_0+\upsilon t$; a subsequent decrease in particle energy results as shown in the Fig. 1. 

The particle acceleration process could be limited by synchrotron radiation losses. In particular, if we consider the synchrotron radiation power $P_{syn}\simeq 2e^4B^2\epsilon^2/(3m_p^4c^7)$ \cite{Rybicki} and require it to balance the particle acceleration rate, $dE/dt\simeq P_{syn}$ the particle energy will be given by
$$\epsilon_{syn}\simeq\left(\frac{dE}{dt}\times\frac{3m_p^4c^7}{2e^4B^2}\right)^{1/2}\simeq 1.2\times 10^{28}\times\frac{10^{-4}G}{B}\times\frac{R_0}{10^{14} cm}\times$$
\begin{equation}
\label{Esyn} 
\times\left(\frac{1000 nm}{\lambda}\right)^{3/4}\times\left(\frac{L}{2\times 10^{42}\; erg s^{-1}}\right)^{1/4}\times\left(1+2\frac{t}{mos}\right)\; eV,
\end{equation}
where $B$ is the magnetic field, which, together with the wave-length and luminosity, are normalized to their respective values at the peak of the particle energy for the $M = 12\times M_s$ scenario (blue curve). For an order-of-magnitude estimate, we assumed that the magnetic field at the stellar surface is of order $\sim 1$ G and that, during the supernova explosion, it decreases with radial distance, $r$, as $B\propto 1/r$ \cite{SN1}. This requires the initial radii of the stars, for which we used the approximate fit to the stellar radius–mass relation for the range $(8-20)\times M_s$ from \cite{torres}, $R_{st}/R_s\simeq 0.49(M/M_s)^{0.96}$ (where $R_{st}$ and $R_s$ are the stellar and solar radii respectively). 

As can be seen from Eq.(\ref{Esyn}), the synchrotron mechanism is important when the energies are of the order of $10^{28}$ eV, which is several orders of magnitude larger than those shown in Fig. 1. This indicates that synchrotron radiation does not impose any significant constraint on the acceleration process.

From a general physical perspective, it is natural to consider whether some other mechanism like inverse Compton scattering could limit the acceleration process. It is well known , however, that inverse Compton cooling is strongly suppressed for protons \cite{ahar}. Thus neither synchrotron losses, nor inverse Compton losses would have any significant effect on the acceleration process proposed in this paper. 

During the acceleration process, relativistic protons interact with the ISM; that will be another channel for cooling. The time scale for ISM cooling can be approximated as (\cite{ahar}):
\begin{equation}
\label{tpp} 
\tau_{pp}\simeq 5.3\times 10^7\times\frac{1\;cm^{-3}}{n_{_{ISM}}}\; yr,
\end{equation}
where $n_{_{ISM}}$ is the number density of ISM. On the other hand, the acceleration timescale
\begin{equation}
\label{tacc} 
\tau_{acc}\simeq\frac{\epsilon_{max}}{dE/dt}\simeq 5\times 10^{-7}\times\frac{\lambda}{1000\; nm}\times\left(\frac{7\times 10^9\; ergs\; s^{-1} cm^{-2}}{F}\right)^{1/2}\; sec,
\end{equation}
where $F$ denotes the flux of EM radiation and (physical quantities have been normalize to their respective values at the peak for the $ = 12\times M_s$ scenario), is many orders of magnitude shorter than the ISM cooling time. Therefore, the proton–proton interactions have negligible effect on the resonant energization process. 

Interestingly, the following formula for the peak energy for the fixed $L_0 = 10^{43}$ erg/s 
\begin{equation}
\label{Efit} 
\epsilon_{peak}\simeq 3.79\times 10^{14}\times\left(\frac{M}{M_s}\right)^{-0.282}\; eV,
\end{equation}
seem to represent the numerical data (illustrated in the graph) rather well in the range (of mass) $(8-20)\times M_s$.

\section{Conclusions}
For the supernova explosion scenario, at its relatively early stage, we considered the acceleration of protons  through a resonant interaction between the electromagnetic wave and the quantum KGP wave associated with a proton; The KGP can gain massive amounts of energy from high intensity EM background. 

We considered typical supernova events of progenitor masses of $(8; 12; 20)\times M_s$, and showed that the resonant acceleration mechanism can accelerate protons  to high energies of $3\times 10^{14}$ eV.

The energy curves as a function of time exhibit a particularly interesting behavior. Specifically, although the luminosity decreases continuously, the resulting particle energies initially increase, reach a maximum, and only then begin to decrease with time. As discussed above, this behavior is a consequence of the specific nature of the resonant acceleration mechanism.
 
\begin{acknowledgments}
The work of ZO was supported by Shota Rustaveli National Science Foundation of Georgia (SRNSFG) Grant: FR-24-1751. ZO acknowledges the Max Planck Institute for Physics (Munich) for its hospitality during the work on this project.
\end{acknowledgments}

\begin{contribution}
All authors contributed equally.

\end{contribution}






\end{document}